\documentclass[12pt,preprint]{aastex}
\usepackage{url}
\usepackage{amsmath}
\usepackage{color}
\usepackage{textcomp}
\usepackage{mathtext}
\usepackage{amssymb}
\usepackage{wasysym}

\newcommand{\ssst}{\scriptscriptstyle}
\newcommand{\E}[1]{\times 10^{#1}}

\newcommand{\RA}[3]{{#1}^{{\rm h}}{#2}^{{\rm m}}{#3}^{\rm s}}
\newcommand{\decl}[3]{{#1}^{\circ}{#2}^{\prime}{#3}''}

      \newcommand{\ps}{\,{\rm s}^{-1}}
    
\newcommand{\cm}{\,{\rm cm}}    \newcommand{\km}{\,{\rm km}}
\newcommand{\kms}{\,{\km\ps}}       
\newcommand{\kpc}{\,{\rm kpc}}
\newcommand{\erg}{\,{\rm erg}}

\newcommand{\MHz}{\,{\rm MHz}}  
\newcommand{\GHz}{\,{\rm GHz}}  
\newcommand{\um}{\,\mu{\rm m}}    

\newcommand{\nHH}{n({\rm H}_{2})} \newcommand{\NHH}{N({\rm H}_{2})}
\newcommand{\VLSR}{V_{\ssst\rm LSR}}
\newcommand{\du}{d_{9}}

\newcommand{\twCO}{$^{12}{\rm CO}$}   
\newcommand{\thCO}{$^{13}{\rm CO}$}
\newcommand{\eiCO}{${\rm C}^{18}{\rm O}$}
\newcommand{\Jotz}{$J$=1-0}    
\newcommand{\Jtto}{$J$=2-1}
\newcommand{\Jttt}{$J$=3-2}

\newcommand{\xray}{{\rm X-ray}}

     \newcommand{\Chandra}{{\sl Chandra}}
  \newcommand{\Fermi}{{\sl Fermi}}

\newcommand{\snr}{{\rm Kes\,73}}  

\newcommand{\gray}{{\rm $\gamma$-ray}}

\begin{document}

\title{A Study of Fermi-LAT GeV $\gamma$-ray Emission towards 
the Magnetar-harboring Supernova Remnant Kesteven 73 and Its Molecular Environment}
\author{Bing Liu\altaffilmark{1},
        Yang Chen\altaffilmark{1,2,6},
        Xiao Zhang\altaffilmark{1},
        Qian-Cheng Liu\altaffilmark{1},
        Ting-Lan He\altaffilmark{1,3},
        Xin Zhou\altaffilmark{4},
        Ping Zhou\altaffilmark{5,1},
        Yang Su\altaffilmark{4}
        }

\altaffiltext{1}{\footnotesize Department of Astronomy, Nanjing University,
163 Xianlin Avenue, Nanjing 210023, China}

\altaffiltext{2}{\footnotesize 
Key Laboratory of Modern Astronomy and Astrophysics,
Nanjing University, Ministry of Education, Nanjing 210093, China}

\altaffiltext{3}{\footnotesize Department of Astronomy, Beijing Normal University,
Beijing 100875, China}

\altaffiltext{4}{\footnotesize Purple Mountain Observatory, 2 West Beijing Road,
Nanjing 210008, China}

\altaffiltext{5}{\footnotesize Anton Pannekoek Institute, University of Amsterdam, PO Box 94249, 1090 GE Amsterdam, The Netherlands}
\altaffiltext{6}{\footnotesize Corresponding author.}

\begin{abstract}

We report our independent GeV \gray\ study
of  the young shell-type supernova remnant (SNR) Kes\,73
 which harbors a central magnetar,
and CO-line millimeter observations toward the SNR.
Using 7.6 years of \Fermi-LAT observation data,
we detected  an extended \gray\ source (``source A") with centroid
on the west of the SNR, with a significance of $21\sigma$
in 0.1--300\,GeV and an error circle of $5.'4$ in angular radius.
The \gray\ spectrum cannot be reproduced  by a pure leptonic
emission
or a pure emission from the magnetar,
and thus a hadronic emission component is needed.
The CO-line observations reveal a molecular cloud (MC) at $\VLSR\sim90\km\ps$,
which demonstrates morphological correspondence with the western boundary
of the SNR brightened in multiwavelength.
The \twCO\ (\Jtto)/\twCO\ (\Jotz) ratio in the left (blue) wing 85--88$\km\ps$
is prominently elevated to $\sim1.1$ along  the northwestern boundary,
providing kinematic evidence of the SNR-MC interaction.
This SNR-MC association yields a kinematic distance $9\kpc$
to Kes\,73.
The MC is shown to be capable of accounting for the hadronic
\gray\ emission component.
The \gray\ spectrum can be interpreted with a pure hadronic emission
or a magnetar+hadronic hybrid emission.
In the case of pure hadronic emission, the spectral index of the
protons  is 2.4,  very similar to that of the  radio-emitting
electrons, essentially consistent with the diffusive shock acceleration  theory.
 In the case of magnetar+hadronic hybrid emission,
a magnetic field decay rate $\ga10^{36}\erg\ps$
is needed to power the magnetar's curvature radiation.

\end{abstract}


\section{Introduction}
\label{sec:intro}
Cosmic rays (CRs) are relativistic particles 
that are mainly comprised of hadrons (protons and nuclei) with a
small fraction of leptons. The origin of CRs remains 
a highly controversial issue as yet, although they have been known for 
more than a hundred years.
During propagation, relativistic hadrons 
may interact with subrelativistic nuclei, 
producing $\pi^0$ mesons that will decay to {\gray}s.
Moreover, highly energetic leptons can produce {\gray}s 
by inverse Compton (IC) scattering low-energy photons or 
by nonthermal bremsstrahlung emission.
Thus, \gray\ observations can offer us crucial information 
to solve the puzzling issue.  
Supernova remnants (SNRs), known for tremendous energy 
contained in their strong shocks, are one of the most popular kinds of 
candidates for  Galactic CR  accelerators.
Dozens of GeV \gray\ sources associated with SNRs 
have been discovered by the  \Fermi\ Large Area Telescope (LAT) in recent years 
\citep[e.g.,][]{Abdo2009W51C,Abdo2010W44SCi,1fsnr}.
However, it is still difficult to distinguish the 
radiation processes of the {\gray}s (hadronic or leptonic)
 that are associated with SNRs,
and it becomes more complicated for the SNRs with  poorly characterized
interstellar environments and associated compact object.
In this paper, we will study the GeV $\gamma$-ray emission from
the magnetar-harboring SNR~Kesteven 73 (hereafter Kes~73)
and its possible molecular environment.

Kes~73 (G27.4$+$0.0) is a young SNR with an incomplete shell both in radio and \xray{s},
filled with clumpy X-ray substructures \citep{Helfand1994kes73,Kumar2014kes73}.
It hosts the magnetar 1E~1841$-$045 
that was first identified as  an  anomalous X-ray pulsar (AXP)  in the center
\citep{Vasisht1997kes73}.
The \xray\ studies  
inferred that its forward shock has encountered the interstellar/circumstellar material
\citep{Kumar2014kes73, Borkowski2017}.
The remnant is estimated to be young, of an age between 500 and 2100 years
\citep{Tian2008kes73, Kumar2014kes73,Gao2016mn,Borkowski2017},
while the characteristic age of the AXP is $\sim4.7$ kyr.
Its progenitor is suggested to be a massive star 
($\gtrsim20$ M$_{\odot}$, \citealt{Kumar2014kes73};  
$<20$ M$_{\odot}$, \citealt{Borkowski2017}).
Moreover,  the distance   to Kes~73 is a controversial issue.
The distance estimated from the H\,{\sc i} absorption is
between $7.5$ and $9.8\kpc$ \citep{Tian2008kes73}. 
Use of a statistical method for pulsar distance measurement
gives the distance to the central magnetar 1E 1841-045
as $8.2$--$10.2\kpc$ \citep{Verbiest2012}. 
In addition, the $\Sigma$-$D$ relationship suggests a distance of 
$\sim14.8\kpc$ \citep{Pavlovic2014}.

Magnetars are a small group of X-ray pulsars 
comprised of AXPs and soft $\gamma$-ray repeaters,
with strong magnetic fields.
Pulsed GeV radiation has been expected to arise from the magnetospheres of magnetars
\citep[e.g.,][]{Cheng2001mag, Beloborodov2007mag, Takata2013mag}. 
As for 1E~1841$-$045, its slow spin period (11.8\,s)
 and rapid spin-down rate imply an extreme field strength,
$\sim 7\times 10^{14}\ \rm G$,
assuming the dipole spin-down model
\citep{Hongjun2013}.
In recent research,  
extended GeV \gray\ emission  
that is possibly associated with the Kes~73 SNR/AXP system was detected,
while the origin of  this diffuse \gray\ emission
remains a controversial issue \citep{1fsnr,Lijian2017magnetar,Yeung2017kes73}. 
In a GeV survey for magnetars, 
no significant detection of  \gray\ flux or \gray\ 
pulsation from 1E 1841$-$045  has been found, 
and the stringent upper limit on the 0.1--10\,GeV emission of  
the magnetar  is $<2.02\times10^{-11}$ erg cm$^{-2}$ s$^{-1}$
after the subtraction of the extended emission from the SNR
\citep{Lijian2017magnetar},
while another \Fermi-LAT study implied that the magnetar is seemingly 
a necessary and sufficient source for the downward-curved spectrum below 10\,GeV
\citep{Yeung2017kes73}.

There are a few hints of dense gases such as molecular clouds (MCs)
in the nearby region.
The 1720\,MHz OH line is detected projectively at $12'$ west of the SNR
at a local standard of rest (LSR) velocity $\VLSR\sim+33\km\ps$
\citep{Green1997}
but  it has been identified with a separate H\,{\sc ii}~region
\citep{Helfand1992}.
Observations in \thCO\ ($J=$1-0) and H\,{\sc i}~lines imply associated
gas at $\sim110\km\ps$ \citep{Tian2008kes73}.
Broadened  features in the  \twCO\ (\Jtto) line  are found from
a  region $\sim2'$--$4'$ away from the SNR boundary and
is conjectured to be caused  by  a disturbance due to the fast-moving
ejecta \citep{Kilpatrick2016co}.

In an independent study, we examine the GeV emission from the Kes~73 region
using \Fermi-LAT data and investigate the interstellar 
molecular environment of the SNR by millimeter CO-line observations toward the region.
We focus on the spectral properties of the $\gamma$-ray emission and
possible hadronic contribution  resulting from the interaction 
between the SNR and the nearby  MC and 
  provide an estimate for the possible contribution from the magnetar.
In the rest of this paper, we describe the $\gamma$-ray and millimeter
observations and data reduction in \S2,
 and present the data analyses and results in \S3.
We discuss the results in \S4 and summarize this work in \S5.

\section{Observations and Data}
\label{sec:ob}

\subsection{\Fermi-LAT Observational Data}
The LAT on board \Fermi\ is a \gray\ imaging instrument 
that detects photons in a broad energy range of 
20 MeV to more than 300 GeV.
Starting from the front of the instrument,
the LAT tracker has 12 layers of thin tungsten converters
(FRONT section), followed by four layers of thick tungsten converters
(BACK section).
Its per-photon angular resolution (point-spread function, PSF,
the $68\%$ containment radius ) varies  strongly with photon energy and 
improves a lot at high energies ($\sim5\degr$ at 100 MeV,
and $0\fdg8$ at 1 GeV, \citealt{Atwood2009fermilat}). 
In addition, the PSF for the FRONT events are approximately a 
factor of two better than the PSF for the BACK events.

In this research, we collected 7.6 years of \Fermi-LAT Pass 8 data, 
from 2008-08-04 15:43:36 (UTC) to 2016-03-25 00:10:13 (UTC). 
We used the package \Fermi\ \emph{ScienceTools}~version {\tt v10r0p5}\footnote{See http://fermi.gsfc.nasa.gov/ssc} released on 2015 June 24, 
to analyze the data in the energy range 0.1--300\,GeV.
We only selected  \emph{Source} (evclass$=$128) events within 
a maximum zenith angle of 90$^{\circ}$ in order to filter out 
the background \gray{s} from the Earth's limb and applied the 
recommended filter string 
``($DATA\_QUAL>0) \&\& (LAT\_CONFIG==1$)" 
in \textit{gtmktime} to choose the good time intervals.
The corresponding instrument response functions (IRFs) 
are ``P8R2\_SOURCE\_V6" for the total (FRONT+BACK) data and 
``P8R2\_SOURCE\_V6::FRONT" for the FRONT data.

\subsection{CO Line Observations and Data}
The observations in millimeter molecular lines
toward SNR Kes~73 were made  in two periods,  
both in position switching mode.
The first observation was made in the  \twCO~(\Jtto) line (at 
230.538\,\GHz) in 2010 January using the K\"{o}lner Observatory 
for Submillimeter Astronomy (KOSMA) 3m submillimeter telescope in Switzerland.
A superconductor-insulator-superconductor (SIS) receiver 
was used as the front end, and 
an acousto-optical spectrometer (AOS) was used as the back end. 
We mapped a $15'\times15'$ area covering Kes~73 centered at
($\RA{18}{41}{17}.3$, $\decl{-4}{56}{17}.0$, J2000) 
with grid spacing $1'$ and the reference position is at  
($\RA{18}{41}{19}.2$, $\decl{-4}{56}{11}.0$, J2000).
The half-power beam width (HPBW) of the telescope  
is $130''$, and the main beam efficiency is $\eta_{mb}\sim68\%$. 

The follow-up observation was made in the \twCO~(\Jotz) line (at 115.271\,\GHz),
the \thCO~(\Jotz) line (at 110.201~\GHz), and the \eiCO~(\Jotz) line (at 109.782\,\GHz)
in 2014 April using the 13.7 m millimeter-wavelength
telescope of the Purple Mountain Observatory at Delingha (hereafter
PMOD), China.  
At the front end, there is a $3\times3$ pixel Superconducting Spectroscopic Array 
receiver, which was made with SIS mixers using the sideband separating scheme 
\citep{Zuo2011Delinha,Shan2012}. 
An instantaneous bandwidth of 1\,\GHz\ was arranged for the back end. 
Each spectrometer provides 16,384 channels with total bandwidth of 1000\,\MHz\ 
and the velocity resolution was $0.158\kms$ and $0.166\kms$ for the \twCO\
and  \thCO\  lines, respectively. 
We mapped a $45'\times45'$ area covering Kes~73 centered at ($27\fdg5,0\fdg0$) 
in the Galactic coordinate system with a grid spacing $\sim 30''$ 
and the reference position is at ($27\fdg5,0\fdg0$).
The HPBW of the telescope is  $52''$ and
the main beam efficiency is $\eta_{mb}\sim52\%$. 

All the CO data were reduced with the GILDAS/CLASS package
\footnote{http://www.iram.fr/IRAMFR/GILDAS} developed by IRAM.
The intensity scales were calibrated using  the standard chopper-wheel 
calibration method \citep[e.g.,][]{Ulich1976} for molecular lines.
Thus the intensity  derived is the one corrected for the atmospheric and ohmic attenuation. 
For extended sources, this intensity needs further correction by the main beam efficiency 
$\eta_{mb}$ to yield an observational radiation temperature.
The mean RMS noise levels of the main beam brightness temperature 
are about 0.5\,K, 0.2\,K, and 0.2\,K for the \twCO~(\Jotz), \thCO~(\Jotz)
and \twCO~(\Jtto)  lines, respectively.

In addition, we also used the \twCO~(\Jttt) data from the second release (R2) data of  
the CO High-Resolution Survey (COHRS) of the James Clerk Maxwell Telescope 
 \citep{Dempsey2013}.

\subsection{Other Data}

In order to compare the distribution of molecular material in the environs of Kes~73
with the morphology of  the SNR, we used the archival \Chandra\ X-ray data (ObsID: 729),
\footnote{http://cda.harvard.edu/chaser/}
the NRAO VLA Sky Survey (NVSS, \citealt{nvss1998}) 
1.4~\GHz\ radio continuum emission data  and
the Wide-field Infrared Survey Explorer (WISE, \citealt{wise2010})  
Band 4 ($22\um$) mid-infrared (IR) all-sky survey data.

\section{Multiwavelength Data Analysis and Results}

\subsection{\Fermi-LAT $\gamma$-rays}
\label{sec:fermi}

Following the standard binned likelihood analysis procedure,
the \Fermi-LAT data analyses were applied to the
$14\arcdeg\times 14\arcdeg$ (in equatorial coordinate system) region of interest, 
which is centered at the position of Kes~73, i.e.,
R.A. = 18$^{\rm h}$41$^{\rm m}$19$^{\rm s}$, 
decl. = $-4^{\circ}55'59''$ (J2000).
The baseline model was generated by 
the user-contributed program {\tt make3FGLxml.py}.
It includes all the Third \Fermi-LAT Catalog (3FGL) sources \citep{3fgl} 
within radius 15$^\circ$ centered at Kes~73 and diffuse background, 
which consists of both the Galactic and extragalactic components 
(as specified in the files \emph{gll\_iem\_v06.fits} and
\emph{iso\_P8R2\_SOURCE\_V6\_v06.txt}, respectively).

\subsubsection{Detection and Localization}
\label{subsec:loc}
\begin{figure}
\centering
\includegraphics[width=0.48\textwidth]{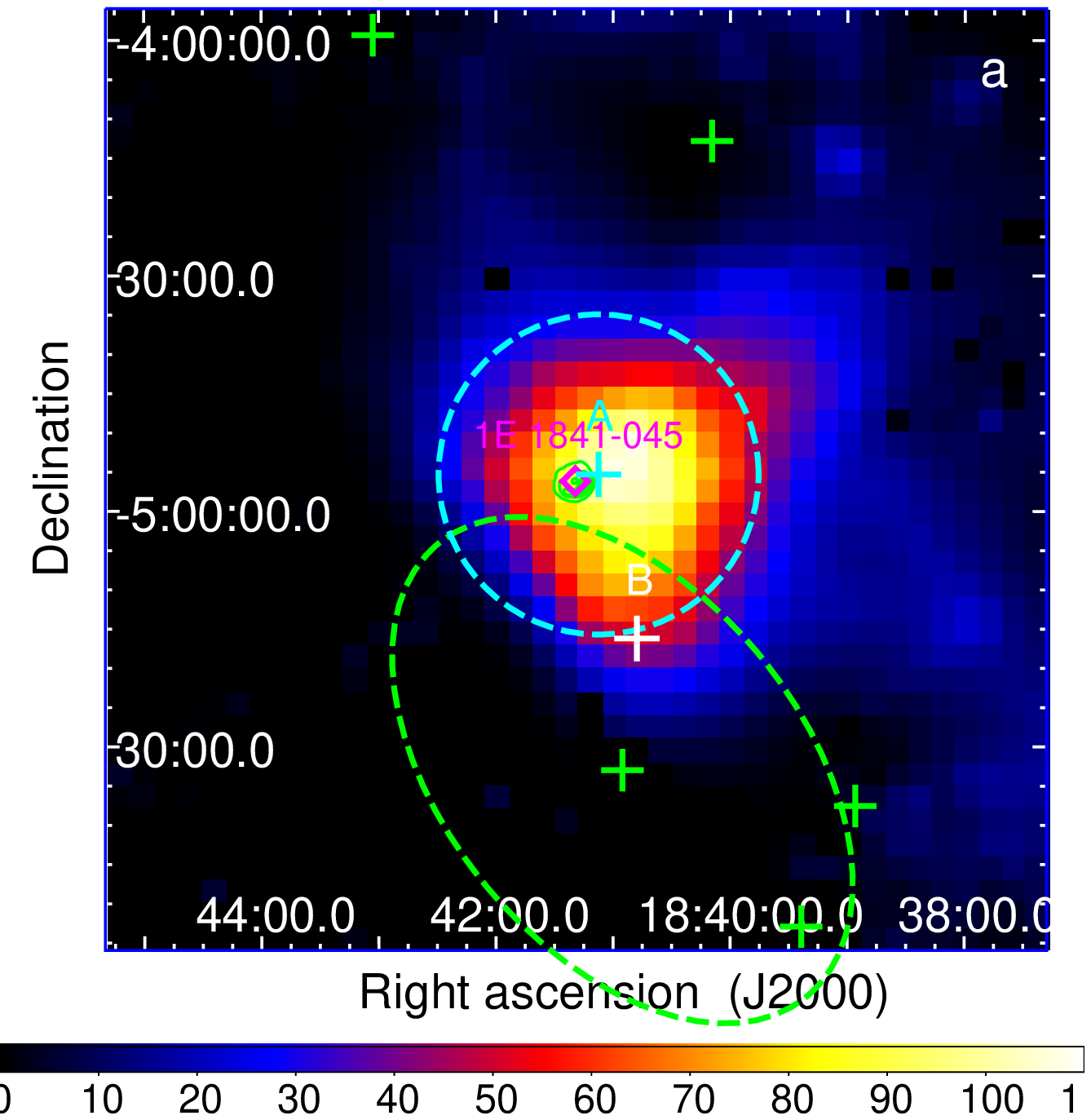}
\includegraphics[width=0.48\textwidth]{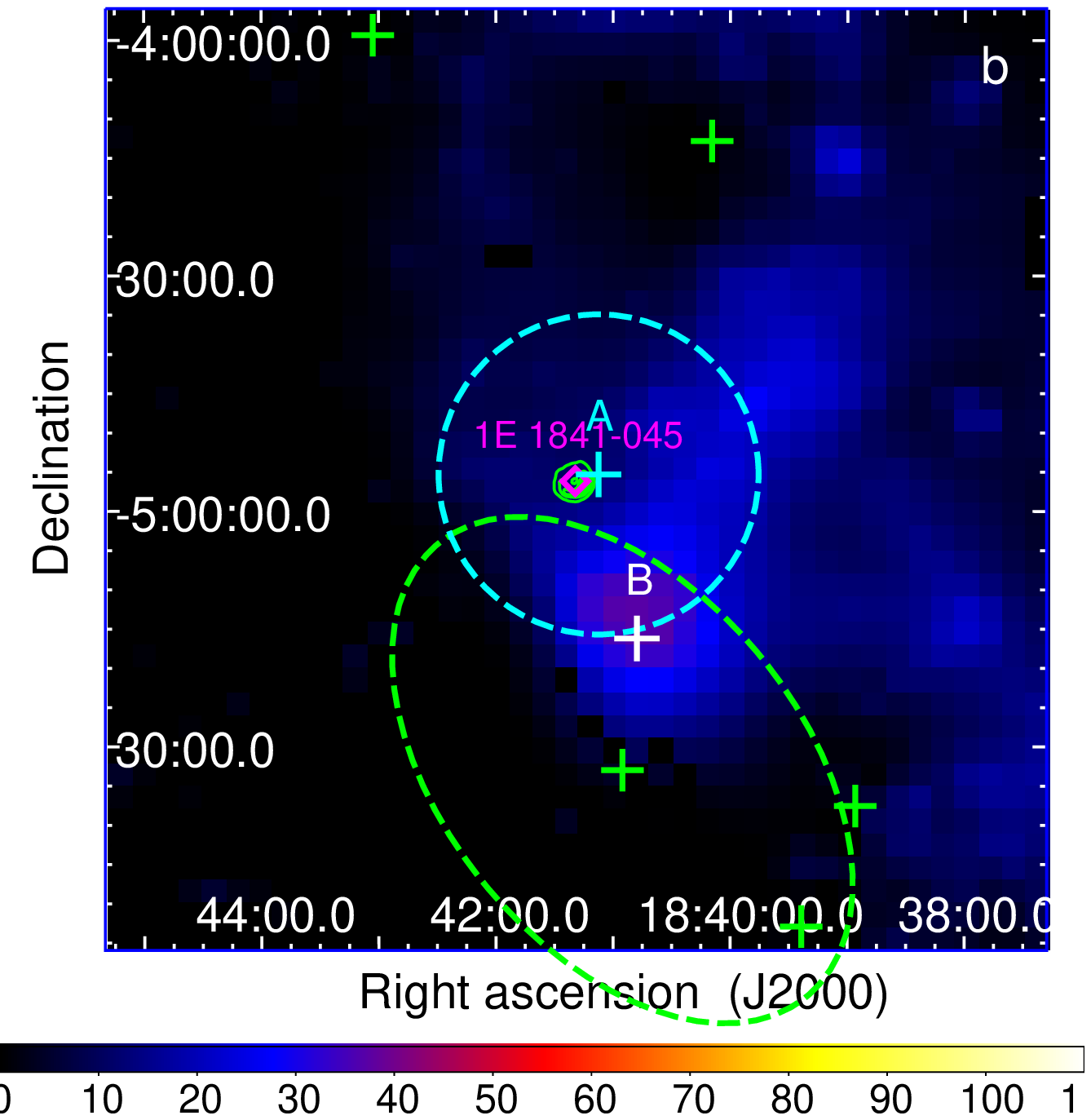}
\includegraphics[width=0.48\textwidth]{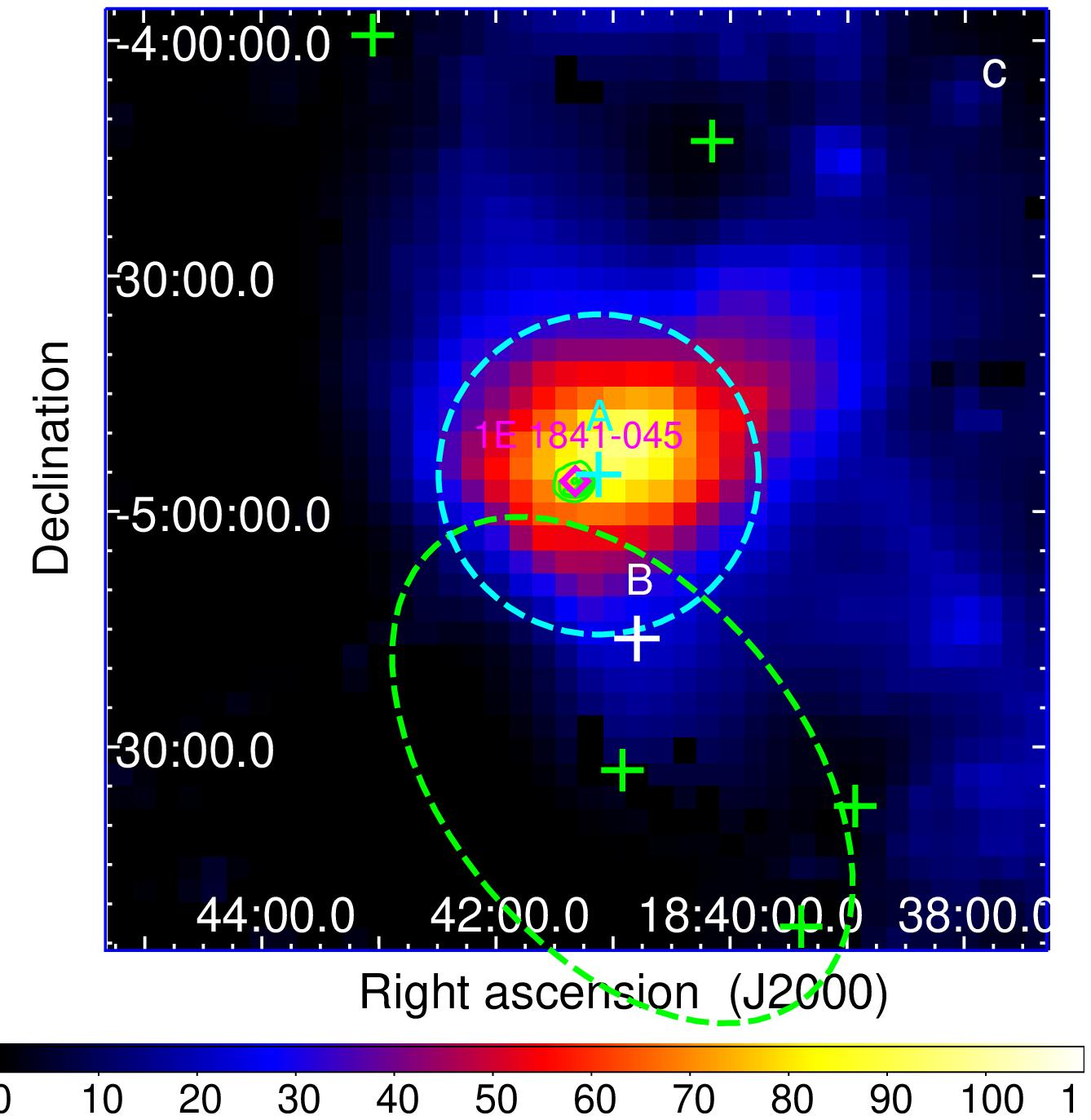}
\includegraphics[width=0.48\textwidth]{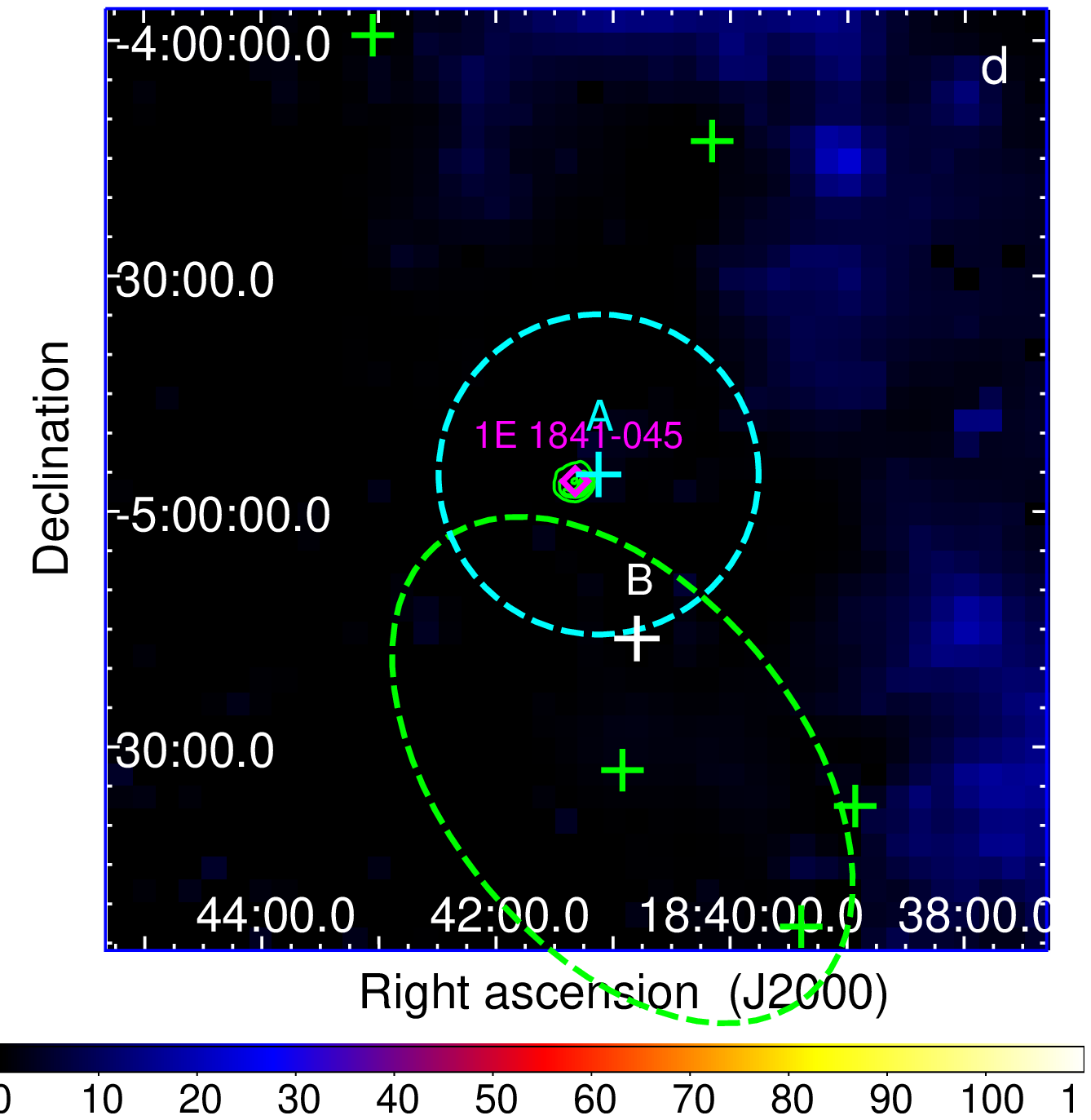}
\caption{TS maps (1--300 GeV) of a $2^{\circ}\times2^{\circ}$ region 
centered at Kes\,73. The image scale of the map is $0\fdg05$ pixel$^{-1}$.
(a) All sources  in the baseline model have been included. 
(b) All sources  in the baseline model 
as well as source~A (as a point source) have been included. 
(c)  All sources  in the baseline model 
as well as source~B have been  included. 
(d)  All sources  in the baseline model as well as source~A 
(as an extended source) and B have been  included.
The green  crosses label the positions of 3FGL sources, 
the cyan cross labels the best-fit position of source A, 
 and its best fitted disk template is shown with a dashed cyan circle.
The white cross  indicates the location of  an additional source (B) 
we added in the source model (for details, see  \S\ref{subsec:loc}).
The  magenta diamond represents the location of 1E 1841$-$045.
The image is overlaid with  NVSS 1.4\,GHz radio  continuum contours.
The  dashed-green oval depicts the spatial template of 
the extended 3FGL  source HESS J1841$-$055.}
\label{fig:tsmap}
\end{figure}

At first,  a binned likelihood analysis (using \textit{gtlike}, 
a tool in the \emph{ScienceTools} package) 
was applied in the energy range 1--300\,GeV.
In order to search for indications of \gray\ emission 
that is probably associated to Kes~73,
we generated a Test Statistic (TS) map for a 
$2^{\circ}\times2^{\circ}$ region centered at Kes~73 (see Figure~\ref{fig:tsmap}). 
The TS value is defined as
 $2(\log{\mathcal L}/{\mathcal L}_0)$, where ${\mathcal L}_0$ is
the likelihood of null hypothesis and ${\mathcal L}$ is
the likelihood with a test source included at a given position.
As shown in Figure~\ref{fig:tsmap}a,
with all components in the baseline model treated as the background,
 the TS value around SNR Kes~73, especially on its western side, was quite high ($\sim110$),
implying lots of residual \gray\ emission in this area.
To approximate the residual \gray\ emission, 
we added a point source with a simple power law (PL) spectrum 
to the baseline model at the TS peak position. 
Then we applied another binned likelihood analysis 
in the energy range 1--300 GeV on the newly updated source model.
Next, we ran \textit{gtfindsrc} (a tool in the \emph{ScienceTools} package) 
and located the best-fit position of the newly added source (hereafter source~A)
at R.A.= 18$^{\rm h}$ 41$^{\rm m}$ 07$^{\rm s}\!$, 
decl.= $-$4$^{\circ}$ 55$'$ 19$''\!$ (J2000)
with a 68\% confidence error circle of $5.'4$ in angular radius.
For the following analyses, the center of source A is 
fixed in this position.

Next, to find out whether such a point source 
could well approximate the residual emission or not,
we generated another TS map centered at Kes~73 
with the contribution from  the newly added source subtracted. 
As can be seen from Figure~\ref{fig:tsmap}b, 
the TS value  to the southwest of Kes~73 was still high ($>25$),
implying many residual {\gray}s that might 
come from another unknown \gray\ source.. 
To test this hypothesis, we performed similar procedures
to those for the detection and localization of source~A.
By adding another point source (hereafter source~B) 
with a PL spectrum at the peak position in the latter TS map,
we adjusted the model and performed another binned likelihood 
analysis in the energy range 1--300 GeV.
Then we located the best-fit position of source B 
at R.A.=$\RA{18}{40}{47}$, decl.= $\decl{-5}{16}{13}$ (J2000) 
with a 68\% confidence error circle of $3'.0$ in angular radius
utilizing \textit{gtfindsrc}.
With the two point sources (A and B) added to the baseline model, 
the residual emission was then found to be negligible and
 the value of the likelihood function increases significantly.
In the energy range 1--300 GeV, the significance 
is  $\sim 10.2\sigma$ for source~A and $\sim 8.8\sigma$ for source B,
 with  both assumed to be point sources.

In the above procedures, adding source~A improves the source model likelihood
by 121.7, and additionally adding source~B further improves the likelihood
by 21.2.
Compared with source~A, which is almost coincident with SNR~Kes~73;
however, 
source~B seems to be far ($21.'6$ projectively) away from the 
SNR.  Therefore, we made another TS map (Figure~\ref{fig:tsmap}c) 
 with source B included in the model as a background source.

\subsubsection{Spatial Analysis}
\label{subsec:spa}

As shown in Figure~\ref{fig:tsmap}c, the residual emission coincident with Kes~73
looks  diffuse rather than point-like.
It was thus necessary to test the extension of source~A.
Using the definition and method in~\citet{Lande2012extend}, 
we modeled the surface brightness profile for an extended source as a uniform disk.
The tested radius range for uniform disk models is $0\fdg2-0\fdg4$ 
with a step of $0\fdg02$.
We performed a likelihood-ratio test
by comparing the likelihood of a uniform disk hypothesis 
(${\mathcal L}_{\rm disk}$) 
with that of point-source hypothesis 
(${\mathcal L}_{\rm point}$)
to test the significance of extension.
The \gray\ source was  considered to be significantly extended 
only if ${\rm TS}_{\rm ext}$, defined as 
$2\log({\mathcal L}_{\rm disk}/{\mathcal L}_{\rm point})$, 
was $\geq16$. 
We listed the ${\rm TS}_{\rm ext}$ values 
obtained from the uniform disk models with 
various radii in Table \ref{tab:ext}.
The highest ${\rm TS}_{\rm ext}$, $40.1$ 
(corresponding to a significance of $\sim 6.3\sigma$)
is achieved when the disk radius is $0\fdg34^{+0\fdg06}_{-0\fdg04}$
\footnote{The $1\sigma$ uncertainties are determined at where the ${\rm TS}_{\rm ext}$ 
is lower than the maximum by 1 according to the $\chi^{2}$ Distribution.}.
As shown in Figure~\ref{fig:tsmap}d, with source~A, source~B,
and other 3FGL background sources included in the background model, 
the residual emission is ignorable. 
Thus, in the following analyses, the \gray\ emission of source~A
is treated as an extended source (of which the centroid appears
on the west of the SNR)  and source~B as a point source, and
the \gray\ emission significances in  the energy range 1--300 GeV for them
are  $\sim 13.3\sigma$ and $\sim 7.3\sigma$, respectively.

\subsubsection{Spectral Analysis}
\label{subsec:spec}

We then performed a spectral analysis for a diagnostic of the
physical property of source~A.
During this process, only FRONT events 
of energy ranging from 0.1 to 300\,GeV were selected 
to lessen the contamination from nearby sources.

We first fit the 0.1--300~GeV spectral data 
of source~A with a PL model. 
Under this assumption, the TS value of source A  is 467.5,
corresponding to a significance of $21.6\sigma$,
and the  obtained spectral shape is relatively flat
with a photon index of $\Gamma=2.21\pm0.06$.
The flux is $(6.11\pm1.14_{\rm stat}\pm1.69_{\rm sys})\E{-11}
\,\mbox{erg}\cm^{-2}\ps$, corresponding to a luminosity
of $\sim5.9\times 10^{35}\du^2$ erg s$^{-1}$, 
where $\du=d/9$kpc is the distance to SNR Kes~73 
in units of a reference value 9\,kpc
(see \S~\ref{S:distance} for details).

Next, we examined other possible spectral models,
such as an exponentially cutoff power law (PLEC), 
a log-parabola model (LogP), and a broken power law (BKPL),
and performed likelihood-ratio tests between the PL model 
(as the null hypothesis) and the other spectral models,
parameterized with index TS$_{\rm model}=2\log({\mathcal L}_{\rm model}/{\mathcal L}_{\rm PL})$. 
The functional forms for these models are
presented in Table \ref{table:form},
while the fitting results were tabulated in Table \ref{table:spec}.
The TS$_{\rm PLEC}$ value $-4.4$
indicates that an additional exponential cutoff (typical for a pulsar) 
does not improve the fitting results compared to the pure PL model.
We find TS$_{\rm LogP}=6.7$ and TS$_{\rm BKPL}=14.4$, with
corresponding significance of $\sim2.6\sigma$ and  $\sim3.8\sigma$, respectively,
indicate a possibility for a curved spectrum 
or a spectral steepening above $\sim1$\,GeV.
However, these two TS values are below the threshold 16,
not high enough for the latter two models
to replace the PL model.

The spectral energy distribution (SED) 
of source~A was extracted via the maximum likelihood analysis of the FRONT events
in seven divided energy bands from 0.1 to 300 GeV (see Table~\ref{tab:flux}). 
The spectral normalization parameters of the sources 
within $5^{\circ}$ of Kes~73 
and the diffuse background components were set free,
while all the other parameters were fixed.
In order to estimate the possible systematic errors 
caused by the imperfection of the Galactic diffuse background model,
we artificially varied the normalization of the Galactic diffuse background 
by $\pm6\%$ from the best-fit values in each energy bin \citep{Abdo2009W51C}.
The maximum deviations of the flux due to these changes in 
the Galactic diffuse background intensities 
were considered as the systematic errors.
The resultant \gray\ spectrum is given in Figure~\ref{fig:sed}.

\subsubsection{Timing analysis}
\label{subsec:tim}
 
To search for long-term variability in source~A,   we follow
 the method described in \citet{2fgl}  and calculate 
its Variability\_{Index} (${\rm TS}_{\rm var}$) 
by dividing  FRONT events in the energy range 0.1$-$300 GeV 
into approximately monthly time bins. 
If the flux is constant,  then ${\rm TS}_{\rm var}$ is distributed as $\chi^2$ with 90 
degree of freedom, 
and variability would be considered probable  if ${\rm TS}_{\rm var}$ exceeds
the threshold 124.1 corresponding to 99$\%$ confidence.
The ${\rm TS}_{\rm var}$ of source~A with all 91 time bins  in 0.1$-$300 GeV is 98.9. 
Therefore, there is no significant long-term variability detected in source~A.

In an attempt to explore the origin of the \gray\ emission of source A, 
we searched in the SIMBAD Astronomical Database \citep{simbad}
within the $3\sigma$ error circle of the source for 
possible candidates of its counterpart(s). 
Among all the known objects (stars, MCs, H\,{\sc ii} region, etc) in this area, 
SNR Kes~73 or the Kes~73 /1E~1841$-$045  system is most likely
to be associated to the  $\gamma$-ray emission 
given the  GeV \gray\ analysis results. 

\subsubsection{Comparison with Previous Research}

The detection of the GeV-bright extended source, ``source A",
through the \Fermi\ data analyses,
confirms the discovery  reported previously
\citep{1fsnr, Lijian2017magnetar, Yeung2017kes73}.
But our analysis has some  differences from theirs.
In \citet{Lijian2017magnetar}, on the assumption that
 the extended \gray\ emission is produced  only by the SNR,
they obtained a stringent  upper limit on the 0.1--10\,GeV emission 
of 1E~1841$-$045 ($<2.02\times10^{-11}$ erg cm$^{-2}$ s$^{-1}$)
via adding a point source at the position of 1E~1841$-$045.
Taking a different approach, \citet{Yeung2017kes73} treated these diffuse {\gray}s 
as a combination of the contribution from SNR  Kes~73 and magnetar 1E 1841$-$045,
and performed spectral analyses of  the source (\emph{Fermi} J1841.1$-$0458 therein)  
in two energy bands (0.1–-10\,GeV and 10–-200\,GeV) separately.

In our analysis, the centroid of the extended source~A
is located  to the west of the SNR, nearly on the western edge of the SNR,
which is offset from  the centroid of \emph{Fermi} J1841.1$-$0.458.
The angular radius of source A is a bit larger than that of 
\emph{Fermi} J1841.1$-$0.458.
Moreover, our spectral analysis was performed on the whole energy band (0.1--300\,GeV)
instead of  dividing them into two energy band  (0.1--10\,GeV and 10--200\,GeV).
Also, we added source B as a background source.
Due to these different treatments, the flux of source A
 that we obtained  in 0.1--10\,GeV is a bit lower than the flux of  
\emph{Fermi} J1841.1$-$0.458  reported by \citet{Yeung2017kes73}.

\subsection{CO Line Emission}
\label{subec:co}

\begin{figure}
\centering
\includegraphics[width=0.49\textwidth]{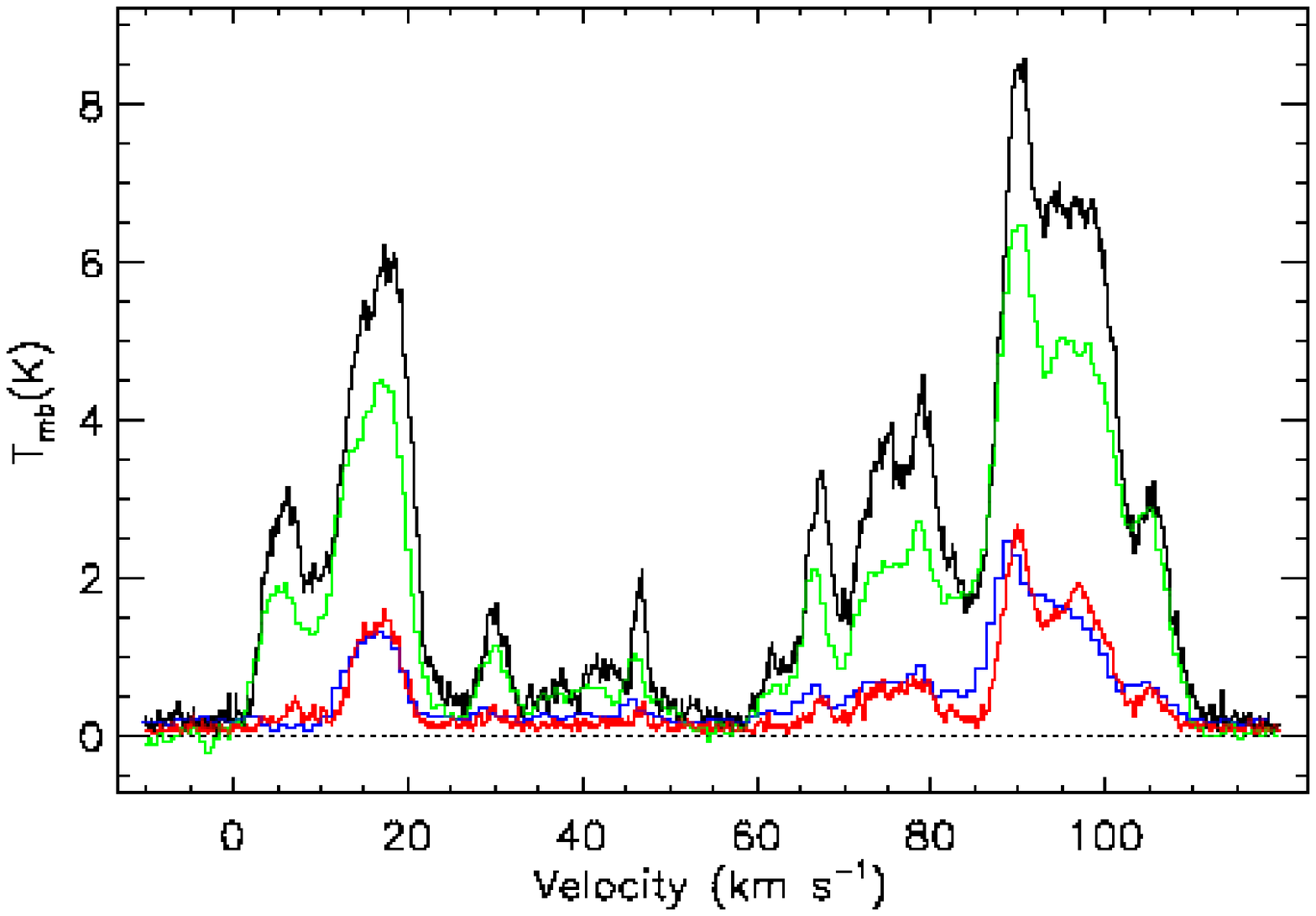} 
\includegraphics[width=0.48\textwidth]{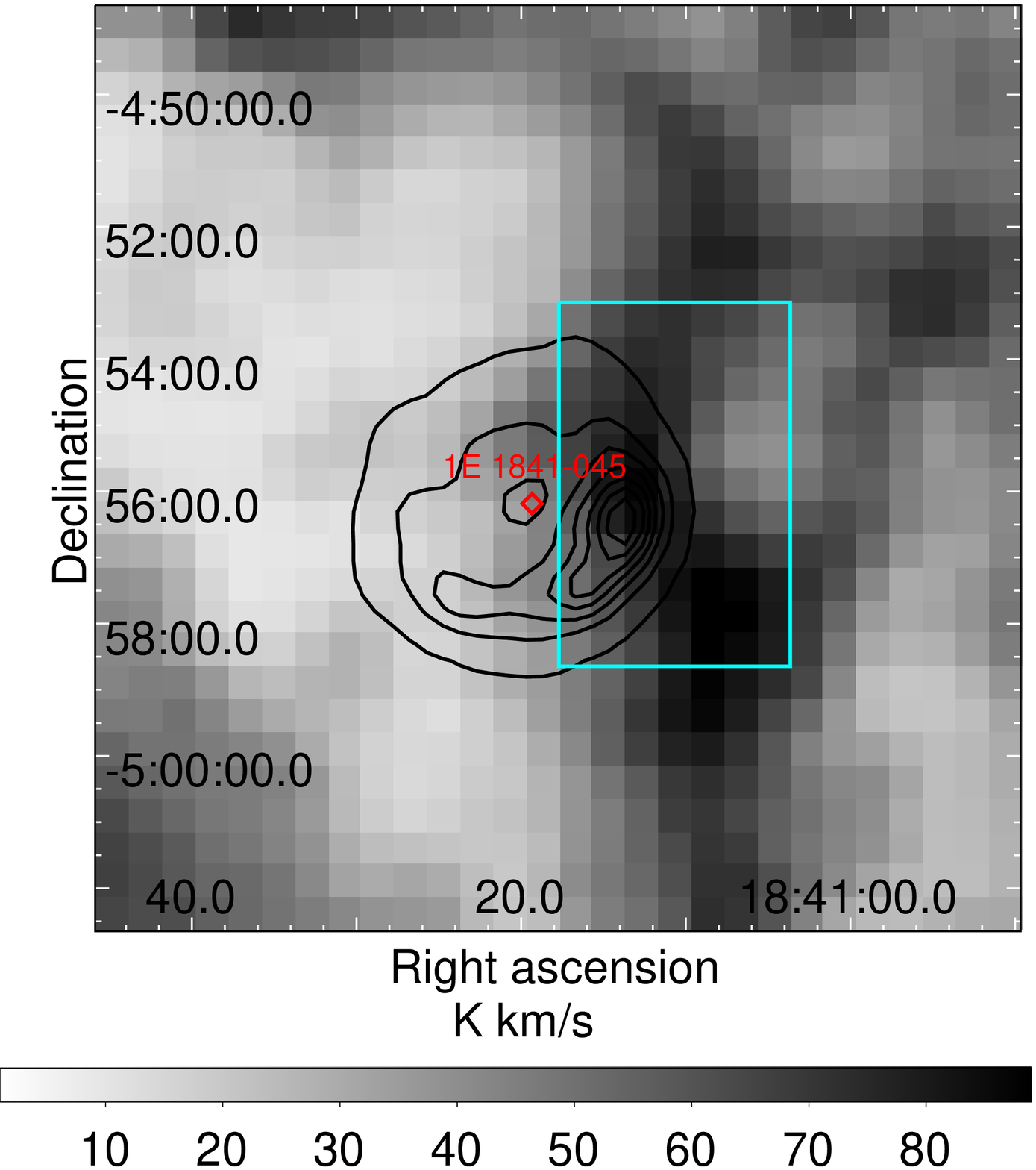}
\caption{{\rm Left:}  average CO spectra from a $3.5'\times 5.5'$  region centered
at R.A.$=18^h41^m10^s.7$, decl.$=-4^\circ55'53''.7$,
covering the SNR Kes\,73,  which has an LSR velocity range of 10--$120\kms$.
The black line is for \twCO~(\Jotz), the green line is for  \twCO~(\Jtto),
the blue line is for \twCO~(\Jttt), and the red line is for \thCO~(\Jotz). 
 {\rm Right:}  \twCO~(\Jotz)  integrated  intensity  map 
in the velocity range   85--$96\kms$.
The map is overlaid with  the NVSS $1.4\GHz$ radio continuum contours
with levels 6, 63, 121, 178, 235, 293, and 350 mJy\,beam$^{-1}$.
The cyan box indicates the region  (region ``W")
from which we extracted  the CO-line spectra.}
\label{fig:co_spec}
\end{figure}

The GeV \gray\ emission (source~A, \S\ref{sec:fermi}) 
that is very likely to be associated with SNR Kes~73
seems to have a centroid on the west of the SNR
and the \gray\ spectrum seems unlikely to be accounted for
with  a SNR leptonic component 
or a magnetar emission alone 
(see \S\ref{subsec:origin} below).
We explore the possible molecular gas contributing to the hadronic component
to the emission.
For this purpose, we extracted CO-line spectra (Figure~\ref{fig:co_spec}, left panel)
from a $3.5'\times5.5'$ region in the western boundary of the SNR (region ``W" defined in
Figure~\ref{fig:co_spec}, right panel).
The spectra are presented in a broad LSR velocity range
$\VLSR=-10$ -- $120\kms$; and
beyond this range, virtually no \twCO\ and \thCO\ emission is detected. 
There are several prominent peaks in the spectra at 
$\VLSR\sim10, 18, 30, 47, 67, 75, 80, 90$, and $110\km\ps$.
By inspecting the intensity maps in the above velocity range, however,
we only found two velocity components (around 18 and $90\km\ps$)
in which the \twCO\ emission demonstrates morphological correspondence with
the SNR.
The possibility of the association of the $\sim18\km\ps$ MC with the SNR can be
precluded due to its improper kinematic distances (see details in \S\ref{S:distance}).
We thus focused on the $90\km\ps$ component.

Figure~\ref{fig:12co10_ch86} and Figure~\ref{fig:12co32_ch86}
show the \twCO\ (\Jotz)- and \twCO~(\Jttt)-line channel maps
in the $86-100\kms$ velocity interval. 
In the interval 85--$96\kms$,  
an extended, curved MC along the north-south orientation appears to overlap
the western region of the SNR Kes~73, 
where the radio, mid-IR and X-ray emission are brightened
(also see Figure~\ref{fig:rgb}).
This demonstrates the morphological agreement between the SNR shell
and the MC.

\begin{figure}
\centering
\includegraphics[width=0.85\textwidth]{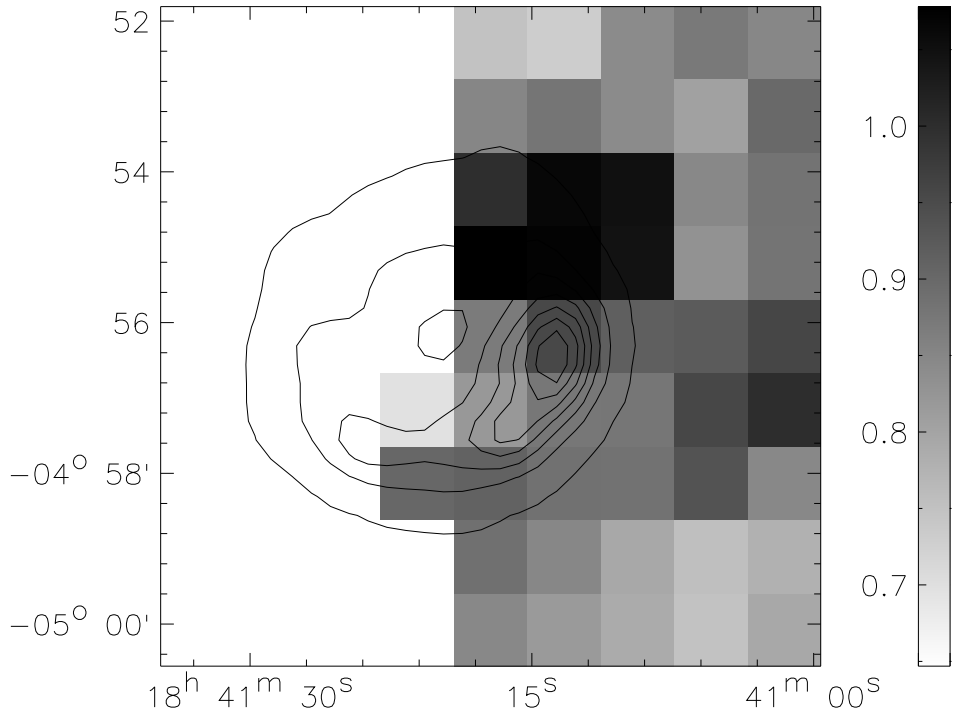}
\caption{ \twCO~\Jtto/\Jotz\ line ratio map  for the LSR velocity range 85--$88\kms$.
 The pixels with \twCO~(\Jotz) or \twCO~(\Jtto) significance $<3\sigma$ are left blank.
The contours are the same as those in Figure~\ref{fig:co_spec}. }
\label{fig:12co_ratio}
\end{figure}

The integrated CO line emission  with a high high-to-low excitation 
line ratio in the line wing is
suggested as a probe of the SNR-MC interaction 
\citep{Seta1998,Jiang20103c397,Chen2014IAU}.
The   \twCO~(\Jtto) /\twCO~(\Jotz) ratio map in the left (blue) wing $85-88\kms$ 
is shown in Figure~\ref{fig:12co_ratio}. 
We see along the northwestern rim the ratios are  prominently elevated to $\sim1.1$.
These locations  with elevated ratios may trace the relatively warm gas
disturbed and heated by the SNR shock, and provide kinematic evidence
for the SNR-MC interaction.

\section{Discussion}

\subsection{The kinematic distance}
\label{S:distance}
It is mentioned in \S\ref{subec:co} that molecular components 
at around 18 and $90\km\ps$ appear to have 
morphological correspondence with the SNR.
Each $\VLSR$ corresponds to two (near and far) kinematic distances.
Here we use the Clemens' (1985)  
rotation curve of the Milky Way \citep{Clemens1985}  together with $R_0=8.0\kpc$ 
\citep{Reid1993} and $V_0=220\km\ps$ to estimate kinematic distances
of the two molecular components.
$\VLSR\sim18\km\ps$  corresponds to  1.1\,kpc or  13.1\,kpc,
but they are both outside the allowed range $7.5$--$9.4$\,kpc estimated from
the H\,{\sc i} absorption data \citep{Tian2008kes73},  and thus it is very unlikely for
this component of molecular gas to be associated with the SNR.
$\VLSR\sim90\km\ps$ corresponds to  5.2\,kpc 
or  9.0\,kpc, and the latter  falls  in the allowed range. 
We have shown evidence  in \S\ref{subec:co}  for the physical association
of the $\sim90\km\ps$ MC with the SNR, and therefore 
we adopt 9.0\,kpc as the distance to the MC/SNR association system.

\subsection{Parameters of molecular gas}

We fit the CO emissions with Gaussian lines for the $\sim90 \km\ps$
molecular gas in region ``W", and the derived parameters,
molecular column density $N(\mbox{H}_2)$,  excitation temperature 
$T_{\rm ex}$, and optical depth of \thCO\,(\Jotz) $\tau$(\thCO),
are summarized in Table~\ref{tab:co}.
Here, the distance to the MC is taken to be 9.0\, kpc.
The column density of H$_2$ and the mass of the molecular gas
are estimated using two methods.
In the first method,
the conversion relation for the molecular
column densities,
$N$(H$_2$)\,$\approx 7\times 10^5 N$~(\thCO) \citep{Frerking1982}
is used under the assumption of local thermodynamic equilibrium
for the molecular gas and optically thick conditions for
the \twCO\ (\Jotz) line.
In the second method,
a value of the CO-to-H$_2$ mass conversion factor (the ``X-factor"),
$N$(H$_2$)/$W$(\twCO),
$1.8\times 10^{20}$\,cm$^{-2}$\,K$^{-1}$\,km$^{-1}$\,s
\citep{Dame2001co} is adopted.

\subsection{The origin of \gray~emission in the region of Kes~73}
\label{subsec:origin}

\begin{figure}
\centering
\includegraphics[width=0.85\textwidth]{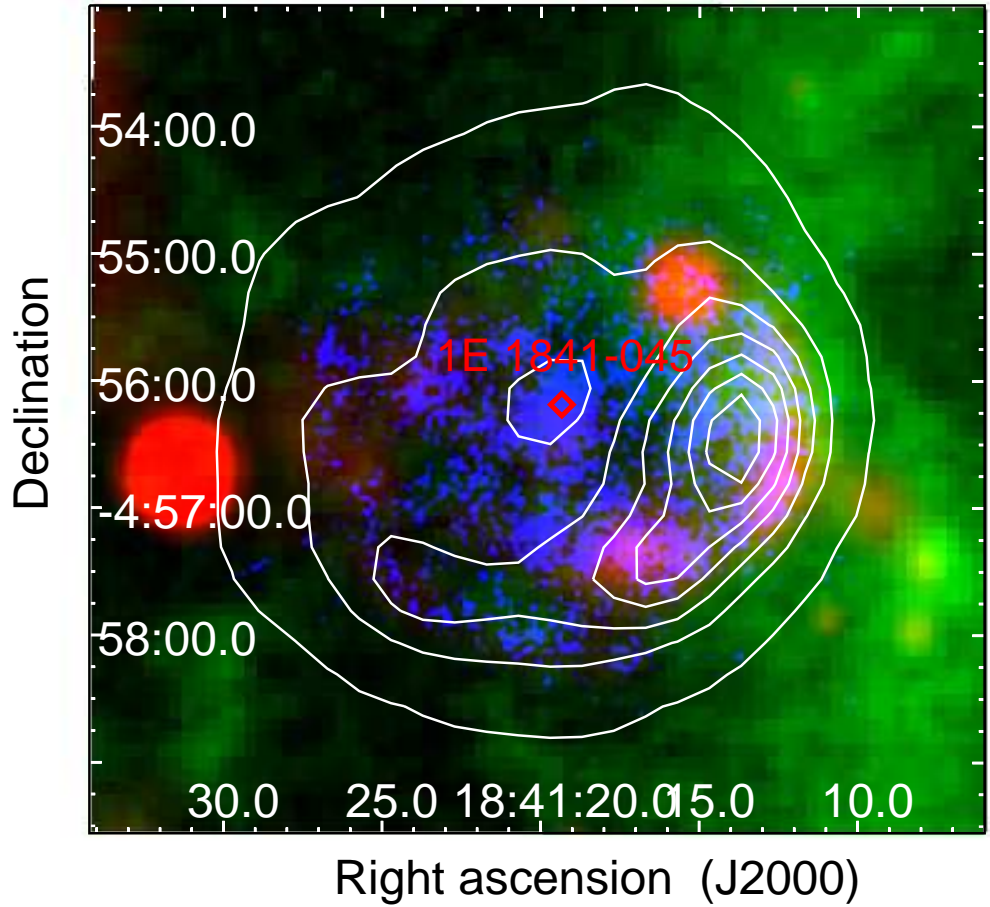}
\caption{Tricolor image of Kes\,73  for multiple wavelengths.
Red: mid-IR emission at $22 \um$  (Band 4)  from WISE observation.
Blue: X-ray  emission (2.7--7\, keV) from \Chandra\ observation (ObsID: 729).
Green: \twCO~(\Jttt)  integrated  intensity map 
 in the velocity range  85--$96\kms$.
The contours are the same as those in Figure~\ref{fig:co_spec}.}
\label{fig:rgb}
\end{figure}

A physical association of SNR Kes~73 with  the $\sim90\kms$ MC 
can naturally explain the enhanced multiwavelength emissions
along the western boundary (see Figure~\ref{fig:rgb}).
The brightened radio emission can result from the magnetic field
compression and amplification when the SNR blast wave 
hits the adjacent western MC and is drastically decelerated.
More dust grains are swept up and heated in the west, which causes
mid-IR enhancement of the western part of the shell.
In a \Chandra\ X-ray analysis of Kes~73 \citep{Kumar2014kes73},
the western boundary (``region 1"  therein) has the highest absorbing
hydrogen column density and volume emission measure of the hard 
component that is ascribed to the forward shock.
This is consistent with the encounter of the forward shock with 
the western dense gas (i.e., MC).
The proximity of the MC on the west of the SNR
is expected to play a role in the hadronic \gray\ emission
from the corresponding region.

\begin{figure}
\centering
\includegraphics[width=0.85\textwidth]{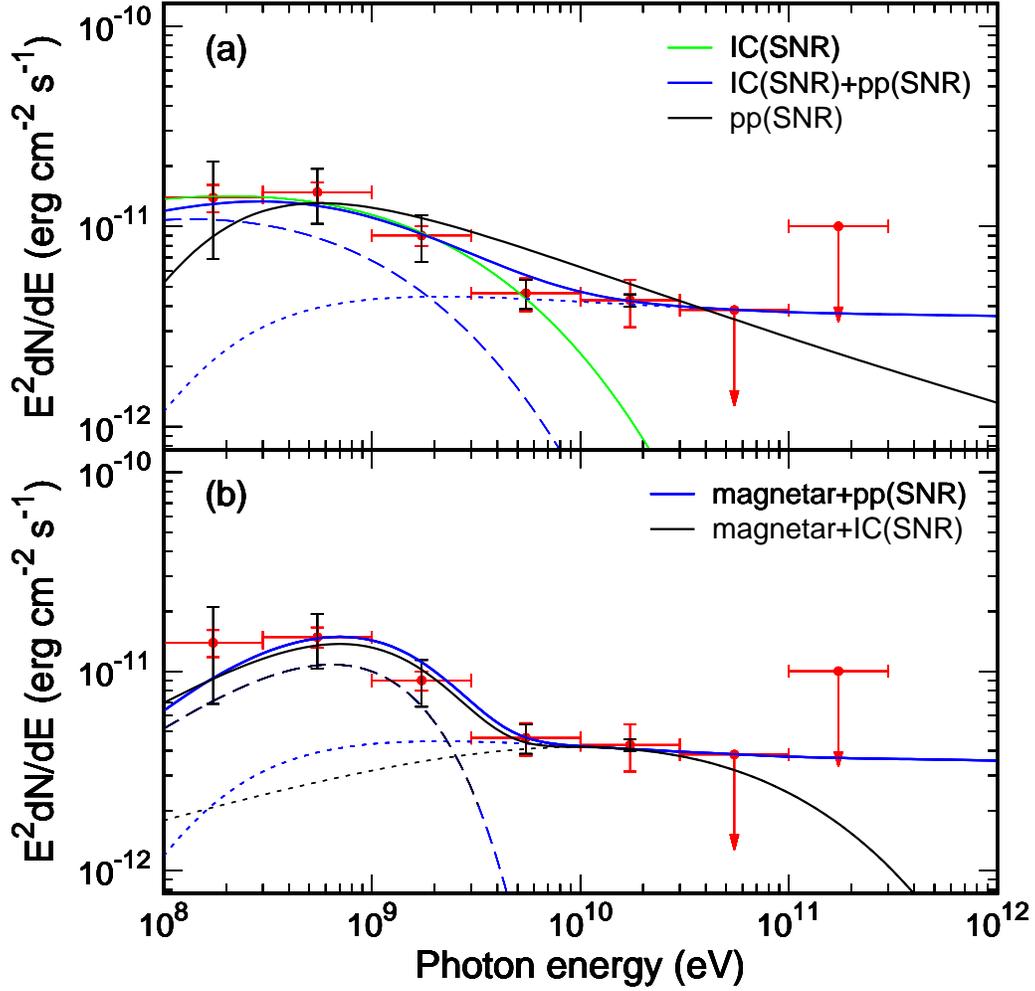}
\caption{\Fermi\ \gray\ SED of Kes\,73, 
fitted with various models (see \S\ref{subsec:origin}).
Systematic errors (see \S\ref{subsec:spec})
are indicated by black bars, and the statistical errors are indicated by red bars.
In the upper panel, the long-dashed and short-dashed lines indicate 
the IC  and p-p interaction components  of Model I, respectively.
In the bottom panel, the long-dashed line represents the magnetar emission
in both Model III and Model IV;
the short-dashed  blue and black lines indicate the p-p interaction  component
in Model III and the IC  component in Model IV, respectively.}
\label{fig:sed}
\end{figure}

We examine the obtained \gray\ spectrum (Figure~\ref{fig:sed})
by analyzing the possible leptonic and hadronic contribution
to the emission from the Kes~73 region.
We assume the accelerated electrons and protons have a PL distribution
 in energy with a high-energy cutoff $E_{i,{\rm cut}}$, namely
\begin{equation}
dN_i/dE_i \propto E^{-\alpha_i} {\rm exp}(-E_i/E_{i,{\rm cut}})
\end{equation}
where $i = e,p$, $E_i$ is the particle kinetic energy, 
$\alpha_i$ is the PL index.  The normalization is determined 
by the total energy in particles with energies above 1 GeV, $W_i$.

We first consider a pure leptonic model in which the \gray{s} 
come from the relativistic electrons scattering off the seed photons, 
e.g., the cosmic microwave background,
but find that the IC \gray{s} from a single population of electrons 
cannot simultaneously account for
the flux data at below and above 10\,GeV.
If only the flux  measurements below 10\,GeV are matched
(see the green line in Figure~\ref{fig:sed}a)
and $\alpha_e= 2.36$ is adopted for a consistency with the radio
emission index 0.68 of the SNR \citep{Green2009snrcatalog},
the fitted parameters are $E_{e,{\rm cut}}=350$\,GeV 
and $W_e=1.6\times10^{51}\du^2$\,erg;
this  energy in electrons is much larger than the energy budget,
a couple of tenths
of the supernova explosion energy ($E_{\rm SN}\sim10^{51}\erg$, canonically),
which  is converted to the accelerated protons and electrons.

So we add a hadronic component 
originating from the decay of $\pi^0$ mesons produced by the pp
interaction between the shock-accelerated protons and the ambient gas. 
In this lepto-hadronic hybrid model
(Model~I), 
we set $\alpha_e=\alpha_p$  (assuming the electrons and protons
are accelerated by the SNR shock),
and $E_{p,{\rm cut}}=3$\,PeV.
The data can be fitted with parameters $\alpha_e=\alpha_p=$2.1, 
$E_{e,{\rm cut}}=200$\,GeV, 
$W_e =$7.8$\times10^{50}\du^2$\,erg,  and
$n_t=17(W_p/10^{50}\ {\rm erg})^{-1}\du^2\,{\rm cm}^{-3}$
where $n_t$ is the average density (averaged over the entire shock surface) 
of the target protons (with which the energetic particles interact),
and $W_p$, the energy in the accelerated protons, is assumed to be
$10\%E_{\rm SN}$.
In Figure~\ref{fig:sed}a, the components of IC emission and
pp emission dominate the flux below and above 2\,GeV, respectively.
In this case, however, the total energy in electrons is still
unreasonably large for an SNR.
Even if we consider the IR photons with energy density
of $1 {\rm eV\,cm}^{-3}$ and spectrum corresponding to a 40\,K temperature,
the energy budget of electrons $W_e$ would be reduced to
 $\sim4.3\times10^{50}\du^2$\,erg, which is still too high.

We  then consider a pure hadronic model (Model~II).
As can been seen from Figure~\ref{fig:sed}a,
the hadronic model seems to be capable of accounting for the \Fermi\ data, too.
Also fixing the cutoff energy as $E_{p,{\rm cut}}=3$ PeV in this case, 
we find $\alpha_p=2.4$ and 
$n_t=40(W_p/10^{50}\ {\rm erg})^{-1}\du^2\,{\rm cm}^{-3}$.
The proton index $\alpha_p=2.4$ is very close to the index 2.36 of the
radio-emitting electrons
(derived from the radio spectral index 0.68, \citealt{Green2009snrcatalog}),
showing the consistency in the frame of the diffusive shock acceleration (DSA) theory.
The proton energy budget means that if the protons accelerated by the SNR shock
take up an energy $\sim10^{50}\erg$,
they can yield the observed \gray{s} by bombarding the proximate
dense gas with an average hydrogen density $n_t\sim 40\ {\rm cm^{-3}}$.

In Model~I, an unreasonable energy in shock-accelerated electrons is
obtained for the IC+pp hybrid case.
The magnetar 1E~1841$-$045 seems to be a possible candidate 
in view of potential energy that could be released from the magnetic field decay
\citep{Takata2013mag,Hongjun2013}.
Indeed, the magnetic field decay rate 
($L_B\ga10^{36}\erg\ps$, \citealt{Yeung2017kes73})
could afford the contribution to the \gray\ emission.
 Therefore, we next explore
 whether the contribution from the magnetar can be  substitute for the IC component,
 namely whether the \gray\ spectrum can be fitted with a combination of
pp emission and emission from  the magnetar (Model~III).
We adopt an outer gap model for isolated pulsars and AXPs
(\citealt{Cheng2001mag,ljzhang2013ApJ}, and specifically eq.(24) therein),
in which the GeV $\gamma$-rays are the curvature radiation released
by the electric field-accelerated electrons/positrons.
Here the energy input is considered to be dominated by the magnetic field decay,
and we assume  $L_B=10^{36}\erg\ps$.
Other model parameters are
the magnetic inclination $\alpha$, the azimuthal angle $\Delta\phi$,
the  dimensionless parameter $\sigma_g$ and  the solid angle of \gray\ beaming
$\Delta\Omega$ (or $f_{\Omega}^{\rm th}=\Delta\Omega/4\pi$).
With the parameters given in Table~\ref{tab:cases},
the flux data can be fitted with the blue solid line in Figure~\ref{fig:sed}b;
the magnetar's emission, represented by the  long-dashed line,
 contributes  to the flux below 10\,GeV, while the pp interaction
 dominates the flux above 10\,GeV.
Like the case of pure IC emission, the magnetar emission alone
cannot reproduce the \Fermi\ \gray\ spectrum.

Next, we explore the case of Model~IV, in which the flux  at low energies
is dominated by the magnetar, while the flux in high energies is
dominated by the IC emission of the SNR shock-accelerated electrons.
The electron index $\alpha_e$ is fixed to 2.36 again.
The model curve (the black solid line in Figure~\ref{fig:sed}b)
appears to also be capable of matching the data points.
However, the energy deposited in the accelerated electrons,
$W_e=1.5\times10^{50}\du^2$\,erg, is again too large to be physical.

\begin{figure}
\centering
\includegraphics[width=0.85\textwidth]{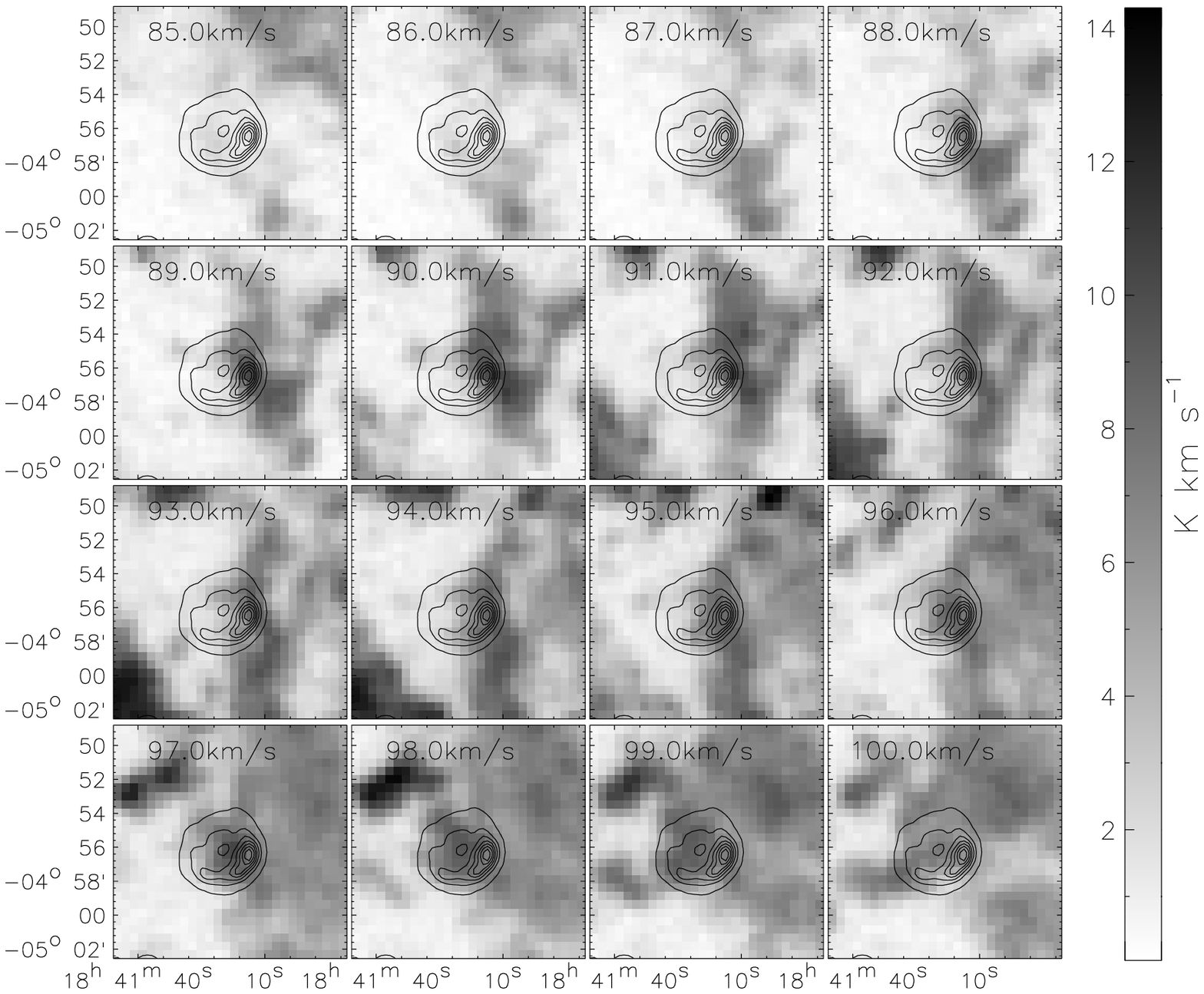} 
\caption{ \twCO~(\Jotz) intensity maps  integrated over successive $1\kms$ 
intervals in the velocity range  
84.5$-$100.5$\kms$.
 The velocity labeling each image is the central velocity of the interval.
The contours are the same as those in Figure~\ref{fig:co_spec}.
 }
\label{fig:12co10_ch86}
\end{figure}

\begin{figure}
\centering
\includegraphics[width=0.85\textwidth]{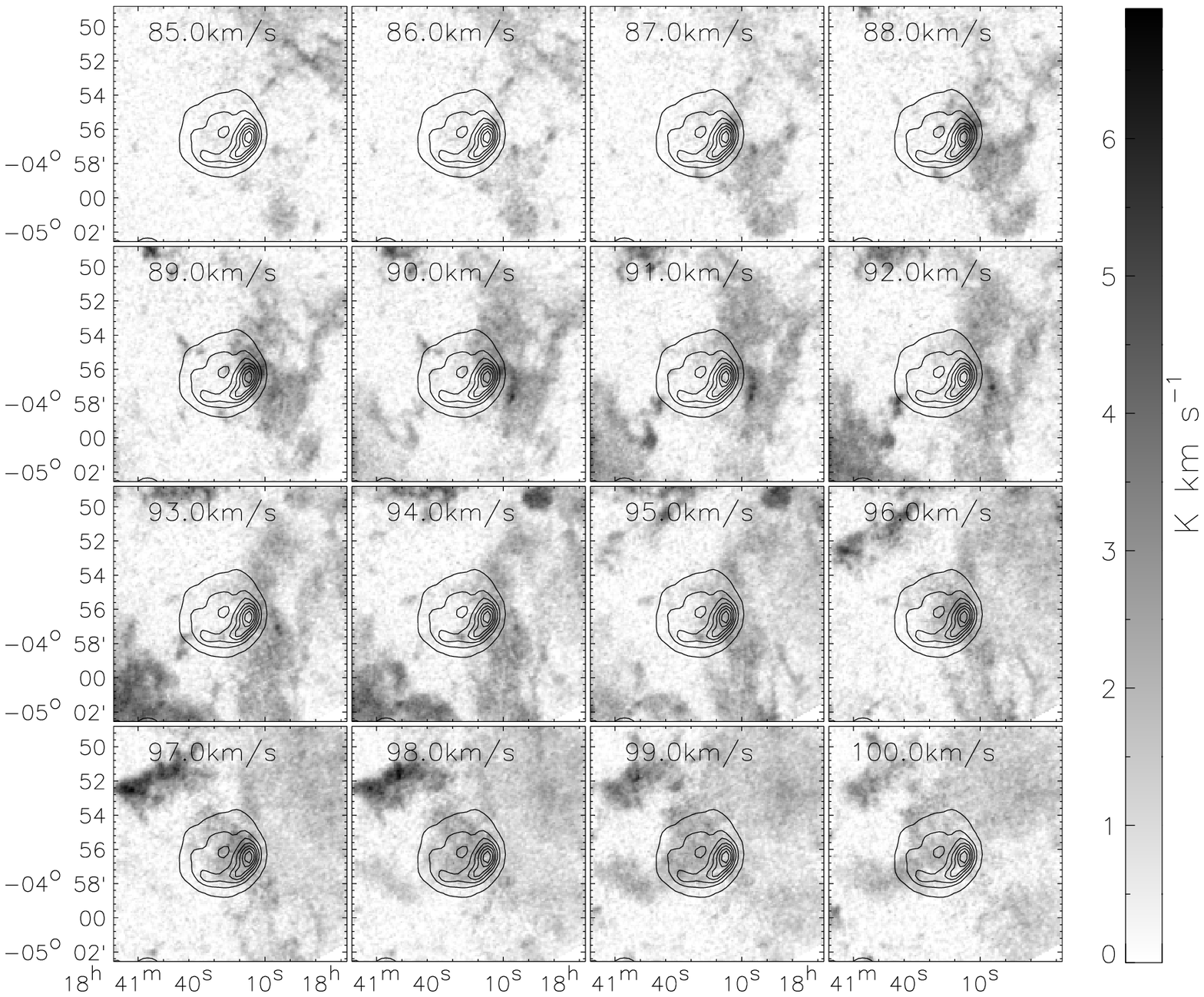} 
\caption{ \twCO~(\Jttt) intensity maps  integrated over successive $1\kms$ 
intervals in the velocity range between 
 84.5 and $100.5\kms$.
  The velocity labeling each image is the central velocity of the interval.
The contours are the same as those in Figure~\ref{fig:co_spec}.}
\label{fig:12co32_ch86}
\end{figure}

In a short summary, the pure SNR IC emission or pure magnetar emission
cannot account for the observed \Fermi\ \gray\ spectrum.
The spectral shape can be reproduced  with  IC+pp or IC+magnetar
hybrid models (Models~I and IV), but the energy  in shock-accelerated
electrons is unphysically large in each case.
Both the pure pp hadronic interaction (Model~II) and
the combination of hadronic and magnetar emissions (Model~III)
seem to be able to account for the spectrum.
They both invoke a dense  adjacent gas
(a few tens of  H atoms cm$^{-3}$).
The $\sim90\km\ps$ cloud appears to be a suitable target of the proton bombardment.
The location of the MC along the western boundary of the SNR
appears to be consistent with the westward offset of the \gray\ centroid
from the SNR center (\S\ref{subsec:spa}).
From Figures~\ref{fig:co_spec}--\ref{fig:rgb},
we crudely estimate the subtended angle of the shock-MC interaction region
as $\sim100^{\circ}$, which corresponds to a solid angle $\Omega^{\rm had}\sim0.7\pi$
or a fraction $f_{\Omega}^{\rm had}\sim1/6$ of the remnant surface.
We obtain an estimate for the density of the molecular gas
$\nHH=(1/2)n_t/f_{\Omega}^{\rm had}\sim120(n_t/40\cm^{-3})\,\cm^{-3}$.
Adopting the column density $\NHH\sim1\E{22}\cm^{-2}$ from Table~\ref{tab:co},
the line-of-sight size of the MC is inferred to be $\sim27(n_t/40\cm^{-3})^{-1}$\,pc.
But the column density given in Table~\ref{tab:co} should be an upper limit
for the associated MC because of possible
contamination from other molecular gas by velocity crowding
in the interval of interest as well as the potential gas at the near distance
(5.2\,kpc, see \S\ref{S:distance}).
Hence the line-of-sight size may be somewhat overestimated.

In Model~III, the central magnetar plays an important role
in the \gray\ emission from the \snr\ region.
Although the timing analysis performed by  \citet{Lijian2017magnetar}
did not detect any statistically significant \gray\ pulsation below 10\,GeV
 being from the magnetar 1E\,1841$-$045,
our \gray\ spectral analysis leaves the possibility of the emission
component from the magnetar.

The extension in our disk model for the \gray\ source
is $0\fdg34^{+0\fdg06}_{-0\fdg04}$ in radius,
similar to
$0\fdg32\pm0\fdg03$ obtained by \citet{Lijian2017magnetar} and
$0\fdg32^{+0\fdg05}_{-0\fdg01}$ obtained by \citet{Yeung2017kes73}.
As noted by \citet{Lijian2017magnetar}, such an extension
is larger than the size of SNR~Kes~73.
The extension is comparable to the size of the  region,
$45'\times45'$,
of our PMOD CO observation. 
The CO emission of the $\sim90\kms$ ($85$--$96\kms$) MC actually pervades
in the region, but the molecular gas is mainly distributed
along the western boundary of the SNR (as shown in
Figures~\ref{fig:co_spec}--\ref{fig:12co32_ch86}).
Apart from the adjacent MC and the magnetar, there may be other
sources that potentially also contribute to the extended emission,
such as a radio complex \citep{Lijian2017magnetar} and
two H\,{\sc ii} clouds (G27.276$+$0.148 and G27.491$+$0.189,
\citealt{Yeung2017kes73}).
Moreover, contamination from the nearby 
background source HESS~J1841$-$055
\citep{Aharonian2008hess}
might be underestimated.
However, given the centroid of the \gray\ emission 
is located on the west of the SNR, 
the detected \Fermi\ GeV \gray\ emission may primarily
arise from the SNR/magnetar system.

\section{Summary}

For the young shell-type SNR~Kes~73 that harbors the central magnetar 1E\,1841$-$045,
we have performed  an independent study of GeV \gray\ emission
and carried out CO-line millimeter observations toward it.
We utilized 7.6 years of \Fermi-LAT observation data in a
$14^{\circ}\times14^{\circ}$ region centered on the SNR.
We find an extended \gray\ source (``source A") with the centroid
on the west of the SNR, with a significance of $21.6\sigma$
in 0.1--300\,GeV and an error circle of $5.'4$ in angular radius.
The \gray\ spectrum cannot be described by a pure leptonic
emission (IC scattering from PL electrons with a cutoff)
 or a pure emission from the magnetar,
and a hadronic emission component seems necessary.
The CO-line observations reveal an MC at $\VLSR\sim90\km\ps$,
which shows a morphological agreement with the western edge of the SNR
that is brightened in multiwavelength.
The ratio between \twCO\ \Jtto\ and \twCO\ \Jotz\ in the left (blue)
wing 85--88$\km\ps$
 is prominently elevated to $\sim1.1$ in the northwestern boundary, 
providing  kinematic evidence of the SNR-MC interaction.
This SNR-MC association yields a kinematic  distance  of $9\kpc$.
It is shown that the MC is an appropriate target for the p-p collision
for generating the hadronic \gray\ emission component.
The \gray\ spectrum can be  interpreted with a pure hadronic emission
or a  magnetar+hadronic hybrid emission.
In the case of pure hadronic emission, the spectral index of the
protons 2.4 is very close to that of the radio
emitting electrons, which is essentially consistent with the DSA theory.
 In the case of magnetar+hadronic hybrid emission,
a magnetic field decay rate $\ga10^{36}\erg\ps$
is needed to power the curvature radiation of the magnetar.
If leptonic emission of the SNR is considered as a component
of the detected \gray{s},
the electron energy budget would be unphysically high.

\begin{acknowledgements}
We are thankful to the staff members of the KOSMA observatory
and PMOD for their support in observations.
This work is supported by the 973 Program grants 2015CB857100 and 2017YFA0402600,
 NSFC grants 11233001, 11633007, 11773014, 11503008, 11590781 and 11403104, and
Jiangsu Provincial Natural Science Foundation grant BK20141044.
This research has made use of the SIMBAD database,
operated at CDS, Strasbourg, France.

\end{acknowledgements}

\bibliographystyle{aasjournal}
\bibliography{cite}
\newpage

\begin{deluxetable}{c|ccccccccc}
\tablecaption{The TS$_{\rm ext}$ values obtained from uniform disk model with various radii}
\tablewidth{0pt}
\startdata
\hline
Radius $r_{disk}$ ($\degr$)      & point & $0.20$  & $0.22$  & $0.24$  & $0.26$  & $0.28$  \\
\hline
TS$_{\rm ext}$           &  ---         &32.0     &33.8      &35.9     & 38.0     & 38.8\\
\hline
\hline
Radius ($\degr$) & $0.30$ & $0.32$  &$0.34$  &$0.36$  &$0.38$  &$0.40$  \\
\hline
TS$_{\rm ext}$         &39.0      &39.6     &40.1      &39.9     &39.6     &38.7 \\
\enddata
\label{tab:ext}
\end{deluxetable}
 
\begin{deluxetable}{c|c}
\tablecaption{Formulae for different spectrum types}
\tablewidth{0pt}
\startdata
\hline
\hline
Name & Formula/Function     \\
\hline  
PL   & $ \mathrm{d}N/\mathrm{d}E = N_0 (E/E_0)^{-\Gamma} $                         \\
PLEC & $ \mathrm{d}N/\mathrm{d}E = N_0 (E/E_0)^{-\Gamma} \exp(-E/E_\mathrm{cut})$  \\
LogP & $ \mathrm{d}N/\mathrm{d}E = N_0 (E/E_0)^{-\Gamma - \beta \log(E/E_0)} $    \\
BKPL &  	$\mathrm{d}N/\mathrm{d}E=\begin{cases} N_0(E/E_\mathrm{b})^{-\Gamma_1} & \mbox{if } E<E_\mathrm{b} \\ N_0(E/E_\mathrm{b})^{-\Gamma_2} & \mbox{otherwise} \end{cases}$\\
\enddata
\label{table:form}
\end{deluxetable}

\begin{deluxetable}{c|cccccc}
\tablecaption{Results from spectral analysis for different spectrum types }
\tablewidth{0pt}
\startdata
\hline
\hline
model  & $\Gamma$ or $\Gamma_1$ & $\beta$ or $\Gamma_2$ & $E_\mathrm{cut}$ or $E_\mathrm{b}$ \,(GeV) &  $2\log({\mathcal L}_{\rm model}/{\mathcal L}_{\rm PL})$ & ${\rm TS}$\tablenotemark{\dagger} \\
\hline
PL     & 2.21$\pm$0.06  & ---            &  ---              &   ---  & 467.5      \\
\hline 
PLEC   & 2.07$\pm$0.13  & ---            & 30.0$\pm$23.6    & $-$4.4   & 420.1 \\  
\hline
LogP   & 1.99$\pm$0.06  & 0.05$\pm$0.02  &  0.3\tablenotemark{\ddagger} &  6.6  & 434.6   \\
\hline
BKPL    &1.79$\pm$ 0.05  & 2.35$\pm$0.02 & 1.00$\pm$0.01     &  14.3  & 429.7  \\   
\enddata
\tablenotetext{\dagger}{TS values of source A  that are obtained from different spectral models}
\tablenotetext{\ddagger}{$E_b$ is a scale parameter that should be set 
near the lower energy range of the spectrum being fit and is usually fixed, 
see \citet{Massaro04-logp}.}
\label{table:spec}
\end{deluxetable}


\begin{deluxetable}{cccc}
\tablecaption{\Fermi\ LAT flux measurements of source~A in the Kes\,73 region}
\tablewidth{0pt}
\startdata
\hline
\hline
$E_{\rm ph}$ (energy band) & 
  $E_{\rm ph}^2dN(E_{\rm ph})/dE_{\rm ph}$\tablenotemark{\dagger} & TS \\
 (GeV)        & ($10^{-12}$ erg cm$^{-2}$ s$^{-1}$) &  \\
\hline
0.173 (0.100--0.300) &    13.9$\pm$2.1$\pm$7.1    &  85.1  \\
0.548 (0.300--1.000) &   14.8$\pm$1.8$\pm$4.5      & 249.6 \\
1.732 (1.000--3.000) &  9.0$\pm$1.0$\pm$2.4      & 136.2  \\
5.477 (3.000--10.00) &  4.6$\pm$0.9$\pm$0.8         &  40.1\\
17.32 (10.00--30.00) &  4.3$\pm$1.1$\pm$0.3       &  20.4  \\
54.77 (30.00--100.0) &  $\le3.8$\tablenotemark{\ddagger} &  1.2 \\
173.2 (100.0--300.0) &  $\le10.0$\tablenotemark{\ddagger} &  5.2 \\
\enddata
\tablenotetext{\dagger}{The first column of errors lists statistical errors and
the second lists systematic errors.}
\tablenotetext{\ddagger}{The 95$\%$ upper limit.}
\label{tab:flux}
\end{deluxetable}

\begin{deluxetable}{l|cc|ccc}
\tablecaption{Model parameters}
\tablewidth{0pt}
\startdata
\hline
\hline
   & &pp& & IC &  \\
\cline{2-6}
 Model & $\alpha_p$ & $n_t$&  $\alpha_e$   & $E_{e,{\rm cut}}$&   $W_e$ \\
    &         &$(W_p/10^{50}\ {\rm erg})^{-1}\du^2\,{\rm cm}^{-3}$ &   & TeV & $10^{50}$\,erg\\
\hline            
I. lepto-hadronic & 2.1 &17     & 2.1  & 0.2 &  7.8\\
II. hadronic & 2.4  & 40 & ---  & --- & --- \\
III. magnetar\tablenotemark{\dagger} + hadronic & 2.1  & 17 &  --- &  --- & ---  \\
IV. magnetar\tablenotemark{\dagger} + leptonic &  ---  & --- & 2.36  &  2.5  & 1.5
\enddata
\tablenotetext{\dagger}{Parameters used for the magnetar are:
$L_B=10^{36}\,erg\ps$,
$\alpha=49\arcdeg$,
$\delta\phi=270\arcdeg$,
$\sigma_g=0.22$, and
$f^{\rm th}_{\Omega}=0.6$ (see text for details).}
\label{tab:cases}
\end{deluxetable}

\begin{deluxetable}{ccccc}
\tabletypesize{\scriptsize}
\tablecaption{Fitted and Derived Parameters for the MCs 
Around $90\km\ps$ in Region ``W"\tablenotemark{a}
\label{parameter}}
\tablewidth{0pt}
\tablehead{
\multicolumn{5}{c}{Gaussian components} \\
\cline{1-5} \\
\colhead{Line} & \colhead{Center ($\km\ps$)} & \colhead{FWHM ($\km\ps$)} & 
\colhead{$T_{\rm peak}$\tablenotemark{b}(K)} & \colhead{$W$(K\,$\km\ps$)}  \\
}
\startdata
\twCO (\Jotz) & $90.1$ & 5.9 & 8.1 & 50.9 \\
\thCO (\Jotz) & $89.9$ & 4.2 & 2.5 & 11.0  \\
\cutinhead{Molecular gas parameters}
$N$(H$_2$)($10^{21}$cm$^{-2}$)\tablenotemark{c} & $M(10^4\du^{-2}M_\odot)$\tablenotemark{c} 
&$T_{\rm ex}$(K)  \tablenotemark{d}& $\tau$(\thCO)\tablenotemark{e} \\
\hline \\
13.5/9.2 & 2.9/2.0 &15.3 & 0.37 
\enddata
\tablenotetext{a}{The region is defined in Figure~\ref{fig:co_spec}.}
\tablenotetext{b}{ Main beam temperature derived from Gaussian fitting of 
CO emission line. }
\tablenotetext{c}{See the text for the two estimating methods.}
\tablenotetext{d}{ The excitation temperature of CO
calculated from the maximum \twCO~(\Jotz) emission point in Region ``W".} 
\tablenotetext{e}{ $\tau$(\thCO) $\approx$ $-$ln[1$-T_{\rm peak}$(\thCO)/$T_{\rm peak}$(\twCO)]}
\label{tab:co}
\end{deluxetable}


\newpage

\end{document}